\documentclass[cits]{PoS}

\title{Maser studies in evolved stars }

\ShortTitle{Maser studies in evolved stars }

\author{\speaker{Francisco Colomer} \\
        Observatorio Astron\'omico Nacional, Calle Alfonso XII 3, 
        E-28014 Madrid, Spain\\
        E-mail: \email{f.colomer@oan.es}}

\abstract{High resolution maps of maser emission provide very detailed
information on processes occurring in circumstellar envelopes of
late-type stars. A particularly detailed picture of the innermost
shells around AGB stars is provided by SiO masers. Considerable
progress is being made to provide astrometrically aligned
multi-transition simultaneous observations of these masers, which are
needed to better constrain the models.  In view of the large amount of
high quality data available, models should now be developed to fully
explain all maser characteristics together (spatial distribution,
variability, etc). New generation instruments (VERA, VSOP-2), new
observational techniques (frequency-phase transfer), and new models
promise important improvements of our knowledge on this topic.}

\FullConference{The 9th European VLBI Network Symposium on The role of VLBI 
in the Golden Age for Radio Astronomy and EVN Users Meeting\\
September 23-26, 2008\\ Bologna, Italy}

\begin{document}

\section{Introduction}
\label{sec1}

The evolution of late-type stars is dominated by their mass-loss
processes. Asymptotic giant branch stars (AGBs) have developed a large
circumstellar envelope (CSE), rich in molecules, which extends to very large
distances (up to 1000 AU from the central star, or 10$^{18}$\,cm). The
expansion of this envelope is caused by radiation pressure on dust
grains, mainly silicates, which form once the temperature and density
have decreased enough (at about 10$^{14}$\,cm).  It is essential to
understand what happens in the innermost layers of the circumstellar
envelope, before dust is formed, to understand the mechanism of mass
loss. Very high resolution mapping of these regions, and of the
envelope in general, is possible thanks to radio observations with
very long baseline interferometry (VLBI). The best probes to study the
physical conditions and dynamics in the different circumstellar shells
in O-rich late-type stars are the maser line emissions of the SiO,
H$_2$O, and OH molecules, which, mainly in the case of SiO, display
very compact structures and high brightness temperatures.

It is well known that AGB stars and envelopes display spherical
symmetry, while later evolution into planetary nebulae (PNe) show
remarkable bipolar structure. The change from AGB to PNe happens
through the proto-planetary nebulae (pPNe) phase, which is very short
(lasts about 1000 years only). Studies of objects in this phase should
provide the information needed to understand the processes involved in
the evolution scenario of these very evolved stars.

\begin{figure}[b]
\includegraphics[width=.95\textwidth]{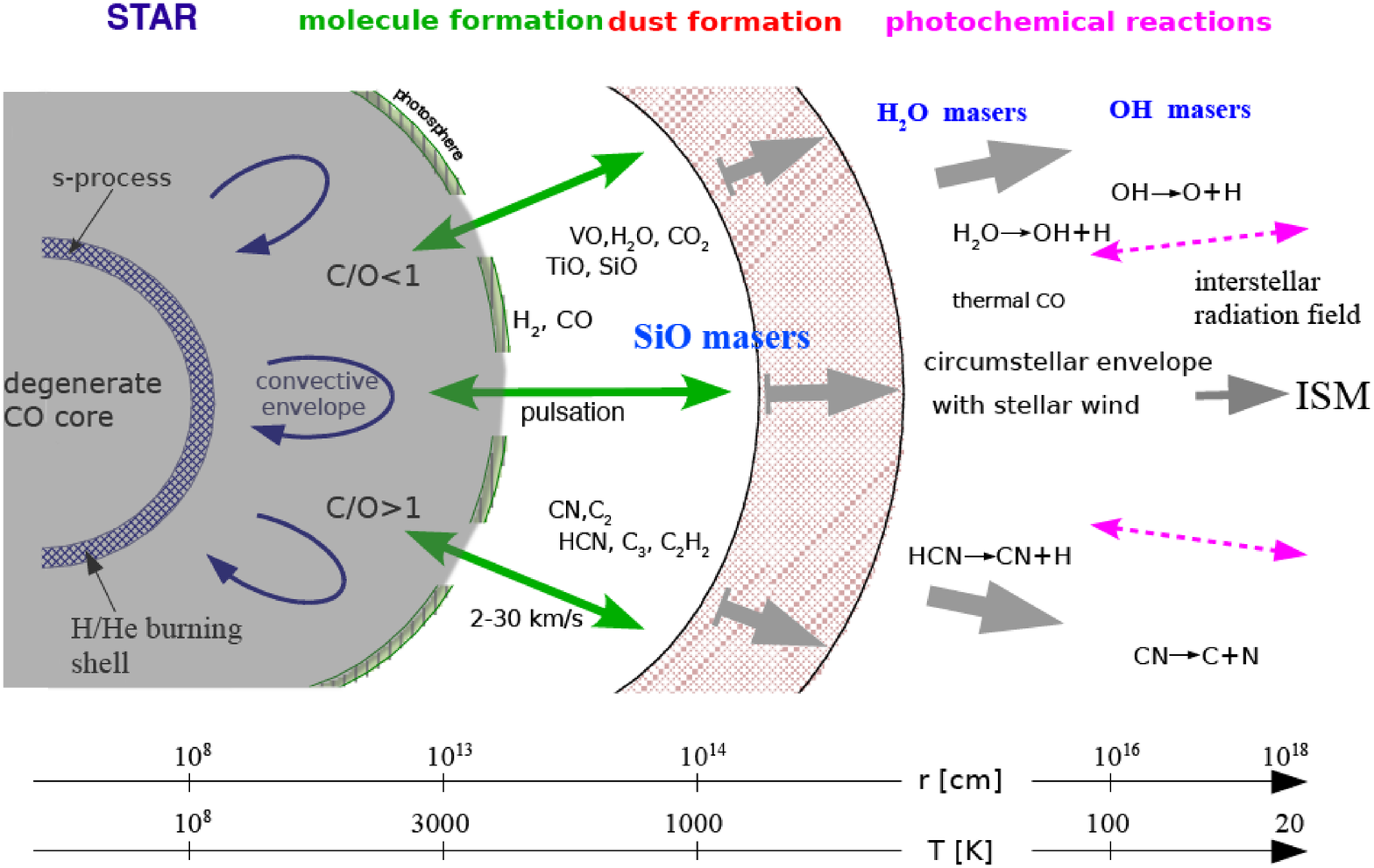}
\caption{Structure of a typical AGB circumstellar envelope \cite{Soria06}. }
\label{cse}
\end{figure}

\newpage

\section{SiO masers in the inner shells of AGB stars}

SiO maser emission is a probe of the physical conditions in the inner
regions of the star circumstellar envelopes. Because interstellar
silicon is incorporated into dust grains (silicates), the dust
formation point marks not only the beginning of the expanding wind but
also the point where most SiO disappears from the gas phase, and so
happens with its masers.

SiO masers have been found mostly in oxygen\--rich late\--type long
period variable stars (ratio $[{\rm O}]/[{\rm C}] > 1$). Positive
detections have been reported in the rotational transitions of many
excited vibrational states ($0 < v \leq 4$), and on isotopes. Very
high spatial resolution observations (using VLBI) have been performed,
most with modern instrumentation \cite{Diamond94} \cite{Greenhill95}
\cite{Desmurs00} \cite{Soria04} \cite{Cotton04} \cite{Yi05}
\cite{Soria05} \cite{Cotton06} \cite{Soria07}. SiO emission appears as
a partial ring structure in most of the objects. The inner envelope
kinematics have been revealed by frequent monitoring with the VLBA
\cite{Gonidakis06} \cite{Gonidakis09}.

\begin{figure}[tb]
\includegraphics[width=.95\textwidth, angle=270]{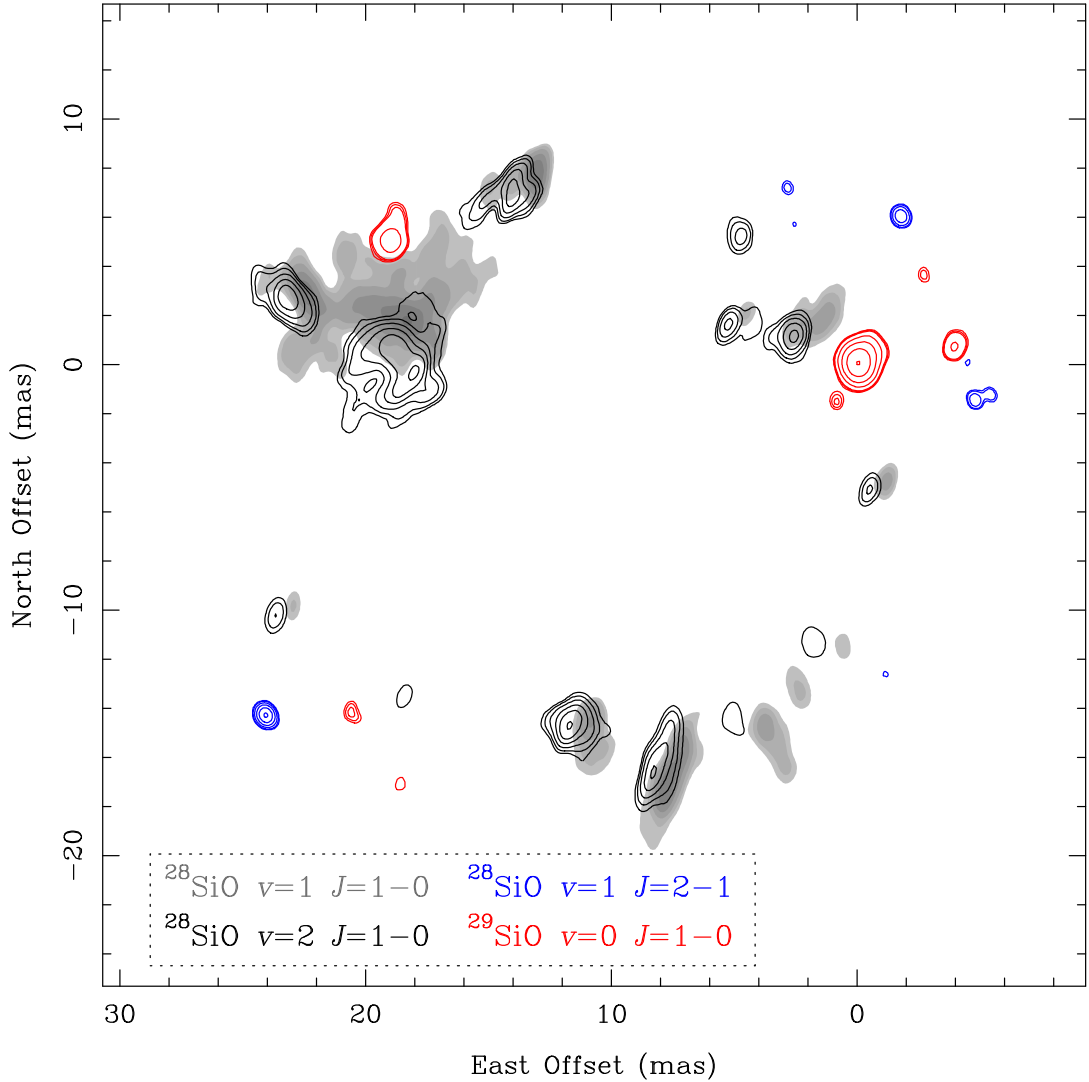}
\caption{Relative distribution of the SiO masers around IRC+10011
\cite{Soria06}}
\label{siomaps}
\end{figure}

The very high quality of the observations mentioned allow to constrain
the models of maser emission. The pumping of SiO masers, for example,
has been described to happen by a radiative and/or collisional
mechanism; these models predict some features that now can be searched
for in the maps. Both expect to find a cascade of masers (i.e.\ the
$J=2\rightarrow 1$ maser overpopulates the $J=1$ state, therefore
favoring the $J=1\rightarrow 0$ maser). Surprisingly, VLBI maps show
that the $v=1$ and $v=2$\ $J=1\rightarrow 0$ distributions are very
similar \cite{Desmurs00} \cite{Cotton06}, in spite of their very
different excitation conditions, while the $v=1$\ $J=2\rightarrow 1$
shows a significantly different spatial distribution \cite{Soria04}
\cite{Soria05} \cite{Soria07}. 
A difficulty which needs to be overcomed is the alignment of
the maps of the different maser lines; strategies have developed from
the alignment of the emission centroid \cite{Desmurs00} to phase
tracking between maser spectral lines \cite{Yi05} and, most
recently, astrometrical alignment (see e.g.~\cite{Rioja08}
\cite{Rioja09}). 
On the other hand, the very weak $v=2$ $J=2\rightarrow 1$ maser
emission in O-rich envelopes was the clue to suggest that the
frequency overlap of different molecular lines could play an important
role \cite{Olofsson81}. Following this idea, it has been possible to
explain both the similar distributions of the $v=1$ and $v=2$
$J=1\rightarrow 0$ lines, and the different behaviour of $v=1$
$J=2\rightarrow 1$ \cite{Soria04}. Figure~\ref{siomaps} is a good
summary of these observational properties, including data on the
ground state maser of $^{\rm 29}SiO$ $v=0$ $J=1\rightarrow 0$.

Some results clearly point in favor of a radiative pumping
mechanism. The SiO maser emission in AGB envelopes is well known to be
time variable, with a period of several hundred days. It correlates
with the IR emission from the central star \cite{Pardo04}. A
collisional mechanism cannot explain this fact; not even the pulsation
of the star travels fast enough to explain this variability
\cite{McIntosh06}. On the other hand, tangential linear polarization
is observed in many of these partial rings \cite{Cotton04}
\cite{Cotton06}, a fact which is naturally explained by a radiative
pumping model \cite{Bujarrabal81}. It is worth noting the remarkably
sophisticated models developed to integrate collisional pumping
mechanisms (see e.g.~\cite{Humphreys96}).

It has been argued that the ring of SiO emission could be used to
estimate the position of the central star, and these being observable
at optical wavelenghts, it would be a tool to relate the radio and
optical celestial reference frames. The direct detection of continuum
radio emission from the stellar photosphere, and the accurate position
of the AGB star, was difficult due to its weakness and large (to VLBI
standards) size. Recent results are very promissing in this aspect
\cite{Reid07}, achieving detection of the o\,Ceti, R\,Leo, and
W\,Hya radio photospheres (with respect to the SiO masers at 43\,GHz)
with the VLA.

\subsection{Astrometry studies with VERA}

The difficulty in achieving astrometrically aligned $v=1$ and $v=2$
SiO maser maps have prevented a unique interpretation of the
observations in terms of the physical underlying conditions, which
depend on the nature of the SiO pumping mechanism. The Japanese {\it
VLBI Exploration of Radio Astrometry}\ (VERA) project is capable to
overcome this limitation by simultaneously observing two sources
(reference calibrator and target) separated almost $3^{\rm o}$ in the
sky. Several AGB stars have been recently investigated in this manner
\cite{Rioja08} \cite{Rioja09} \cite{Kamohara08} \cite{Matsumoto08}. 
The newly developed method of frequency-phase transfer calibration has
been very useful, in some cases, to relate maser emission of different
frequencies.

Thanks to the determination of absolute positions for the SiO maser
spots, VERA also provides measurements of the parallax and proper
motion \cite{Kamohara08}.

\subsection{Results of ultra-high resolution studies}

Ultimate spatial resolution is provided by the global VLBI arrays
operating at high frequencies, as well as on space VLBI baselines. To
date, the Global Millimeter VLBI Array (GMVA) operating at 86\,GHz
matches the angular resolution that is expected from the VSOP-2
project at 43\,GHz, to be launched in 2013. Preliminary detection of
ultracompact SiO maser emission with the GMVA \cite{Colomer09}
indicates that the use of VSOP-2 for these studies is possible. A Key
Science Program is being set for the VSOP-2 mission.

\section{The outer shells: H$_2$O and OH masers}

Because H$_2$O and OH masers have a more extended emission and they
are very well observed with connected interferometry, studies with VLBI
are more scarce. 

The VERA instrument is capable to study also the H$_2$O masers, and
indeed this is its most important design characteristics in order to
study our galaxy's structure and rotation by accurate parallax
measurements. Mapping of these masers in the envelope of AGB stars
have provided some very interesting results \cite{Nakagawa08}, for
example, a precise determination of the distance to VY\,CMa
\cite{Choi08} \cite{Choi09} which in turn permits to better know
its luminosity. The structure and 3-D kinematics of the supergiant
star is also revealed, showing a bipolar outflow along the line of
sight. Parallax and proper motion for S\,Crt are also studied in this way 
\cite{Nakagawa08}.

A more complete picture of the AGB envelope is obtained when also the
OH maser emission is included. Combination of MERLIN, EVN/global VLBI
data permits to compare the emission of SiO, H$_2$O, and OH masers
\cite{Richards07} \cite{Richards09}. The scenario described in
Section \ref{sec1} is challenged, for example in VX\,Sgr, as mainline
OH masers are found interleaved with H$_2$O maser emission. Hints for
such overlap are also found in other stars. New studies with coming
facilities such as e-MERLIN and ALMA will help imaging the star, trace
the dust as it forms, etc.

Observations using absolute phase referencing of the OH maser peaks,
which are expected to coincide with the stellar position, have also
been used to measure the absolute coordinates and parallaxes of
several AGB stars \cite{Vlemmings07}.

\section{Post-AGB objects}

\begin{figure}[bt]
\includegraphics[width=.95\textwidth]{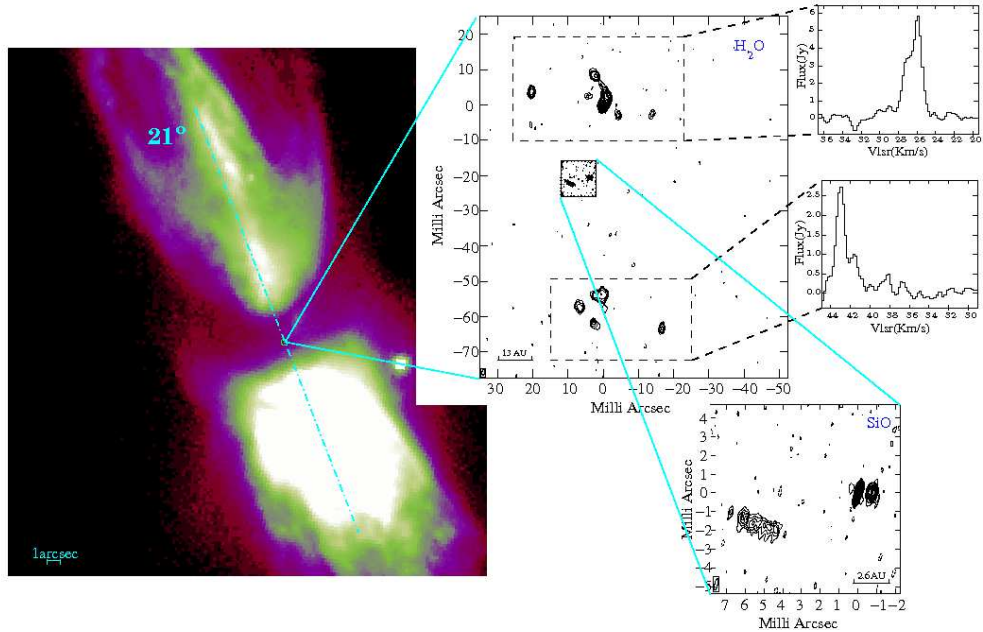}
\caption{Masers at the heart of pPNe OH231.8+4.2 \cite{Desmurs07}.}
\label{oh231}
\end{figure}

It is well known that, while AGB envelopes are in general quite
spherically symmetric, Planetary Nebulae display very different (often
beautiful) bipolar structure. The clue for such change must lie in the
short-lived protoplanetary nebulae (pPN) phase. Masers are not often
observed in pPNe; they have been detected in OH231.8+0.4 with VLBI
(\cite{Desmurs07}). Absolute positions of H$_2$O and SiO masers have
allowed to locate them with great accuracy, also with respect to other
components of the nebulae. While the H$_2$O emission in this object
comes from a region typical of these masers in AGB stars, the bipolar
spatial distribution is peculiar, with two spots and SiO masers in the
middle (see Fig.~\ref{oh231}). The lack of water in the equatorial
regions of the inner circumstellar envelope may be explained by the
presence of a companion. MERLIN and VLTI observations of this object
complement this scenario \cite{Etoka09}.

The study of H$_2$O and OH masers in pPNe is very interesing in a
particular kind of objects, the ``water fountains'', sources showing
maser spikes at very high velocities. The maser emission in them seems
to trace shocks, and proper motions have been measured, see
e.g. \cite{Claussen09}, and references therein.

\section{Summary}

The availability of modern instrumentation has made possible, in the
last years, a great advance in the study of maser emission in evolved
stars. Nowadays, maps of many SiO, H$_2$O, and OH maser transitions
have been produced with very high angular resolution. In many cases,
not only the spatial distribution but also the kinematics and
polarization of the emission has been revealed.

The position of the emission of the different molecular lines relative
to each other is crucial, as it allows to constrain the models in a
way that has never happened before. In this respect, new instruments
that help to provide absolute astrometry (such as VERA) and new
observational techniques (such as frequency-phase transfer) look very
promissing to overcome the traditional limitations.

Having so much data available, and with so high quality, it is time to
devote special efforts to develop models of maser emission whose
predictions fit with the observed maps. Issues such as frequency line
overlap should be included in new generation codes. The future looks
promissing for the field of maser studies in evolved stars.

\vspace{1cm}

\noindent
{\bf Acknowledgements}

I wish to express my gratitude to Valent{\'\i}n Bujarrabal for
providing helpful comments to this manuscript.

\end{document}